\documentclass[pla,twocolumn,showpacs]{revtex4-1}
\usepackage{amsfonts}
\usepackage{amsmath}
\usepackage{amssymb}
\usepackage{graphicx}
\usepackage{epsfig}

\usepackage[unicode]{hyperref}
\hypersetup{
	pdftitle={Vortexlines},
	pdfauthor={Mateo, Yu, Nian},
	pdfsubject={Vortexlines},
	colorlinks=true,
	linkcolor=blue,
	citecolor=blue,
	filecolor=black,
	urlcolor=blue
}

\def\url#1{}

\usepackage{bm}
\usepackage{color}

\usepackage[normalem]{ulem}

\newcommand{\be}{\begin{equation}}

\newcommand{\ee}{\end{equation}}

\newcommand{\bea}{\begin{eqnarray}}

\newcommand{\eea}{\end{eqnarray}}

\newcommand{\nn}{\nonumber }

\begin{document}

\title{Vortex lines attached to dark solitons in Bose-Einstein condensates and Boson-Vortex Duality in 3+1 Dimensions}
\author{A. Mu\~{n}oz Mateo$^{1,3}$}
\email{amunoz@fqa.ub.edu}
\author{ Xiaoquan Yu$^{2,3}$}
\email{xiaoquan.yu@otago.ac.nz} 
\author{Jun Nian$^{4,5}$}
\email{nian@ihes.fr} 
\affiliation{$^1$ Departament de F\'isica Qu\`antica i Astrof\'isica, Universitat de Barcelona, Mart\'i i Franqu\`es, 1, 
	E--08028 Barcelona, Spain }
\affiliation{$^2$ Department of Physics, Centre for Quantum Science, and Dodd-Walls Centre for Photonic and Quantum Technologies, University of Otago, Dunedin, New Zealand}
\affiliation{$^3$New Zealand Institute for Advanced Study, Centre for 
	Theoretical Chemistry and Physics,Massey University, Auckland 0745, New Zealand}
\affiliation{$^4$Institut des Hautes \'Etudes Scientifiques, Le Bois-Marie, 35 
	route de Chartres, 91440 Bures-sur-Yvette, France}
\affiliation{$^5$C. N. Yang Institute for Theoretical Physics, Stony Brook 
	University, Stony Brook, NY 11794-3840, United States}

\begin{abstract}
We demonstrate the existence of stationary states composed of vortex lines 
attached to planar dark solitons in scalar Bose-Einstein condensates. 
Dynamically stable states of this type are found at low values of the 
chemical 
potential in channeled condensates, where the long-wavelength instability of 
dark solitons is prevented. In oblate, harmonic traps, U-shaped vortex lines
attached by both ends to a single planar soliton are shown to be long-lived 
states. Our results are 
reported for parameters typical of current experiments, and open up a way to 
explore the interplay of different topological structures.
These configurations provide Dirichlet boundary 
conditions for vortex lines and thereby mimic open strings attached 
to D-branes in string theory. We show that these similarities can be formally 
established by mapping the Gross-Pitaevskii theory into a dual effective 
string theory for open strings via a boson-vortex duality in 3+1 
dimensions. Combining a one-form gauge field living on the soliton plane 
which couples to the endpoints of vortex lines and a two-form gauge field 
which couples to vortex lines, we obtain a gauge-invariant dual action of open 
vortex lines with their endpoints attached to dark solitons.

\end{abstract}

\maketitle

\section{Introduction}

Quantum vortices and planar dark solitons are frequently generated in current 
experiments with ultracold gases, from Bose-Einstein condensates (BECs) 
\cite{Fetter2009,Frantzeskakis2010} to
Fermi gases \cite{Zwierlein2005,Ku2014,Ku2016}. 
With phase imprinting techniques, the phase of the superfluid can be 
arranged to show either continuous 
$2\pi$ changes around vortex lines, or sudden leaps 
across the soliton planes. 
These topological defects, which separate regions with 
different values of the resulting 
order parameter, also appear after the quench crossing a second-order phase
transition breaking the underlying continuous symmetry known as the Kibble-Zurek 
mechanism~\cite{Kibble1976, Zurek1985}. 
The great achievement of techniques in experiments with ultracold gases has 
directed researchers' attention towards more complex physical systems presenting
nontrivial topologies, which might simulate different types of 
topological excitations discussed in quantum field theory and string theory.  
For instance, analogues of cosmic strings in superfluids
~\cite{Zurek1985,Lamporesi2013} and analogues of Dirac monopoles in the spin-1 
BEC~\cite{Savage2003,Borgh2012} have been proposed. 
${^3}$He A-B interfaces and the boundary surface of the two-component BEC have
been suggested as analogues of branes in string theory~\cite{Bradley2007, 
Nitta-1,Nitta-2, Nitta-3}. Isolated monopoles~\cite{monopolesinBECs} and Dirac 
monopoles \cite{DiracmonopoleBECs} have been observed in recent spin-1 BEC
experiments.

In this work we show that scalar BECs are also suitable for studying rich 
structures of topological defects. In spite of the fact that there have been 
an intensive research about vortices and solitons in scalar BECs, as far as we 
know, the study of topological structures showing junctions between them 
has not been performed. Only a particular configuration of this type has been 
addressed in the context of an effective low-dimensional model in trapped
condensates~\cite{Mateo2013}. Note that in a trapped BEC without
dark solitons stable vortex lines which do not form loops have to end at the 
condensate boundary. Here, within a mean field approach, we address the 
dynamics of scalar BECs 
containing vortex lines that extend between soliton stripes. This 
arrangement makes it possible to find the ends of a vortex line in the bulk of 
the system, just at the position of the soliton. Configurations of this type 
will be referred to as open vortex lines, and they provide Dirichlet boundary 
conditions for the vortex ends. It is interesting to note 
that they mimic open strings attached to D-branes in string theory, in
particular the so-called BIons. Although the arrangements shown in the present 
work are not 
generic from the string theory point of view, they might still offer a ground 
for simulating particular aspects of it. To establish this connection on more 
formal terms, we show that the Gross-Pitaevskii (GP) theory can be mapped
into an effective string theory via the so-called 
boson-vortex duality.

The rest of the paper is structured as follows. In Section 
\ref{sect:model}, we present the mean field model based on the
Gross-Pitaevskii energy functional, which is used to describe accurately the 
dynamics of BECs containing vortices and solitons. In Section 
\ref{sect:dual}, we propose the boson-vortex duality for open vortex 
lines ending on dark solitons. The following sections are dedicated to 
particular 
realistic configurations fitting into both the GP description and its dual
description.  First, in Section 
\ref{sect:openvortex}, we consider condensates trapped along 
their transverse directions and containing axisymmetric vortex lines separated 
by dark solitons that extend over the cross section of the system. We 
demonstrate that a 3D condensate endowed with this 
structure can be a dynamically stable stationary state. Beyond a 
threshold value for the interaction energy such configuration becomes unstable 
due to the soliton decay, and the whole system can evolve through  
new emerging vortex lines.
Second, in Section \ref{sect:halfring}, we study slightly oblate 
condensates holding symmetric planar 
solitons, to which U-shaped vortex lines, that are referred to as half 
vortex rings, are attached. In this variant configuration, each curved 
vortex line has both ends on a single dark soliton. 
We show that, although such stationary state is dynamically unstable,
it has a long lifetime.  The ultimate decay of the soliton shows  chaotic 
behaviors of vortex lines, which might be related to quantum turbulence 
processes. A common feature of both configurations is the presence of 
vortex line pairs separated by the soliton plane. We finish with the 
conclusions in Section \ref{sect:conclusions}, where we emphasize that the 
topological states considered in this paper are feasible to be realized under 
current experimental techniques.

\section{Gross-Pitaevskii Model}
\label{sect:model}
We follow a mean-field approach for modeling dilute gases of repulsive 
interacting bosons making Bose-Einstein condensates at zero temperature. The 
state of such a system is described by a complex order parameter 
$\Psi(\mathbf{r},t)$, whose dynamics is determined by the non-relativistic 
GP Lagrangian density:
\begin{align}
{\cal L}_{GP} = \frac{i\hbar}{2} \left(\Psi\frac{\partial\Psi^*}{\partial t}-\Psi^*\frac{\partial\Psi}{\partial t}\right)-H_{\rm GP} ,
\label{eq:Lagrangian}
\end{align}
where the energy density is given by
\begin{align}
H_{\rm GP} =  
\frac{\hbar^2}{2m}|\nabla \Psi|^{2} + 
\frac{g}{2}\left(|\Psi|^{2}-|\Psi_0|^{2}\right)^2.
\label{eq:free_energy}
\end{align}
Here $g=4\pi\hbar^{2}a/m$ is the interaction parameter, characterized by the 
$s$-wave scattering length $a$ and the atomic mass $m$.
Note that the energy is measured relative to the ground state 
$\Psi_0(\mathbf{r},t)=\psi_0(\mathbf{r})\,\exp(-i\mu \,t/\hbar)$ with chemical 
potential $\mu$. For homogeneous systems, $\Psi_0$ is a real
constant, apart from a global phase, and
$\mu$ fixes the 
value of the uniform density 
$|\Psi_0|^2=\mu/g$. In the presence of an external potential $V(\mathbf{r})$, a 
local chemical potential $\mu_l(\mathbf{r})=\mu-V(\mathbf{r})$ can be defined, 
which in the limit of high interaction energy, or Thomas-Fermi regime,
allows to approximate the non-uniform ground-state density by 
$|\Psi_0(\mathbf{r})|^2=\mu_l(\mathbf{r})/g$. 
In this way, we will use Eq.~(\ref{eq:Lagrangian}) as a unified model for both
homogeneous and non-homogeneous systems. In the latter case, this means 
that when the system is closed, the energy is measured with respect to the ground 
state having the same chemical potential, and thus, in general, a different 
number of particles. For non-homogeneous systems, we will consider 
harmonic trapping potentials with cylindrical symmetry 
$V(\mathbf{r})=V_z(z)+V_\perp(r_\perp)=m\omega_z^2 z^2/2+m\omega_{\bot}^2 
r_\perp^2/2$, having an aspect ratio $\lambda=\omega_z/\omega_\perp$, where 
the transverse coordinates are 
$(\theta,r_\perp)=(\tan^{-1}(y/x),\sqrt{x^2+y^2})$. As a length unit, we 
define the characteristic length of the transverse trap 
$a_\perp=\sqrt{\hbar/m\omega_\perp}$.

The equation of motion corresponding to the Lagrangian density ${\cal L}_{GP}$ 
is the time-dependent GP equation
\begin{align}
i\hbar \frac{\partial\Psi}{\partial t}=\left(  -\frac{\hbar^{2}}{2m}\nabla
^{2} + g( |\Psi|^{2}-|\Psi_0|^2)\right)  \Psi .
\label{eq:tgpe}
\end{align}
Its stationary solutions with chemical potential $\mu$, that is 
$\Psi(\mathbf{r},t)=\psi(z,\theta,r_\perp)$, fulfill the time-independent GP 
equation
\begin{align}
\left(-\frac{\hbar^{2}}{2m}\nabla
^{2} + g( |\psi|^{2}-|\psi_0|^2)\right) \psi = 0.
\label{eq:tgpe1}
\end{align}

\section{Duality}
\label{sect:dual}
\subsection{Effective string action}
\label{sect:action}
The boson-vortex duality is a powerful tool to study the dynamics
of vortex rings~\cite{Lund1976,ZeePaper,Gubser,Gubser16,Horn2015}. It has been 
demonstrated that the dual description of the GP theory in 3+1 dimensions is 
equivalent to a certain type of effective string theory~\cite{ZeePaper, Franz, 
Gubser}. In this paper we extend the dual mapping to open vortex lines and 
argue that dark solitons in the original GP theory could play the role of 
D-branes in the effective string theory by
introducing a pinned boundary condition for vortex lines \cite{BECstring1}. In the 
following, for simplicity we set the trapping potential to be
zero, i.e. $V(\mathbf r)=0$.

Introducing the Madelung transformation  $\Psi = \sqrt{\rho} \, e^{i \phi}$, the equivalent expression of the GP Lagrangian density Eq.~\eqref{eq:Lagrangian} in terms of the condensate density $\rho$  and the phase $\phi$ reads
\be\label{Lag}
\mathcal{L}_{GP}\simeq - \hbar \rho \dot{\phi} - \rho \frac{\hbar^2}{2m} (\nabla \phi)^2 - \frac{g}{2} (\rho - \rho_0)^2\, ,
\ee
where we neglect the quantum pressure term  $\sim (\nabla \sqrt{\rho})^2$,
which should be valid at long wavelengths (the hydrodynamic limit) that we are 
interested in. We rewrite the quadratic term $- \rho 
{\hbar^2}(\nabla\phi)^2/{2m} $  via the so-called Hubbard-Stratanovich 
transformation \cite{HS}, and obtain the dual representation of the original 
theory Eq.~\eqref{Lag} :
\bea\label{LagStar}
\mathcal{L}^{*}_{GP} & \simeq& - \hbar \rho \dot{\phi}- \hbar \nabla \phi   \cdot \mathbf{f} +\frac{m}{2 \rho} \mathbf{f} \cdot \mathbf{f}  -\frac{g}{2}(\rho-\rho_0)^2 \nn\\ 
&=&- \hbar f^{\mu}\partial_{\mu}\phi+\frac{m}{2 \rho} \mathbf{f} \cdot \mathbf{f}  -\frac{g}{2}(\rho-\rho_0)^2,
\eea
where $\partial_{\mu}\equiv (\partial_{t},\nabla)$, and $\mathbf{f}$ is an 
auxiliary three-vector, which is also the space-like components of the four-vector 
$f^\mu \equiv (\rho , \mathbf{f})$. The physical meaning of the 
field $\mathbf{f}$ becomes clear by 
evaluating its equation of motion $\mathbf{f}=\rho\frac{\hbar}{m}\nabla 
\phi$, which matches the superfluid current $\mathbf{J}$ in the GP 
theory described by Eq.~(\ref{eq:Lagrangian})~\cite{Lund1976}.

Let us now decompose the phase $\phi$ into a non-singular part
and a few singular parts related to the topological defects:

\bea \label{phasedecom} \phi=\phi_g+\phi_{\rm cv}+\phi_{\rm ov}+\phi_{\rm s}, \eea 
where $\phi_g$ is the smooth phase, which is the 
Goldstone mode associated with the $U(1)$ symmetry breaking, $\phi_{\rm cv}$ is the
phase of vortex rings, $\phi_{\rm ov}$ is the phase of open vortex lines, 
and $\phi_{\rm s}$ represents the phase of dark solitons.
Note that Eq. \eqref{phasedecom} just means that $\phi_{\rm cv}$, $\phi_{\rm ov}$ and $\phi_{\rm s}$ have singularities,  and they may still contain smooth parts.
The action of the non-singular part $\phi_g$ reads
\bea \label{actionsmooth} \mathcal{S}_{g}&=&\int d^4 x  \left(-\hbar 
f^{\mu}\partial_{\mu}\phi_g\right) \\
&=&\hbar\int d^4 x  
\left[\partial_\mu f^{\mu} \phi_{g}-\partial_{\mu} (f^{\mu}\phi_g)  \right]. \nn 
\eea 
By integrating out the smooth phase $\phi_g$ in the bulk, we obtain
\bea \partial_t \rho+ \nabla \cdot \mathbf{f}=\partial_{\mu} f^{\mu}=0, \eea
which can be solved by
\begin{equation}
f^\mu =\frac{1}{2}\epsilon^{\mu\nu\lambda\sigma} 
\partial_{\nu}B_{\lambda \sigma} \, ,
\label{eq:Bfield}
\end{equation}
where $\epsilon^{\mu\nu\lambda\sigma}$ is the 
totally-antisymmetric tensor, and $B_{\lambda
	\sigma}$ is an antisymmetric rank-2 gauge field. Note that $f^{\mu}$ is invariant under the gauge transformation $B_{\mu\nu} \rightarrow B_{\mu\nu} + \partial_\mu \Lambda_\nu - \partial_\nu \Lambda_\mu $, where $\Lambda_\nu$ is an arbitrary four vector.

In the dual description, the action of a vortex ring in 3+1 dimensions has 
already been 
proposed \cite{Lund1976, ZeePaper, Franz, Gubser}. In this paper we 
focus on open vortex lines. 
For an open vortex line the action reads
\bea \label{actionov} \mathcal{S}_{\rm ov}&=&\int d^4 x  \left(-\hbar 
f^{\mu}\partial_{\mu}\phi_{\rm ov}\right) \\
&=&-\frac{\hbar}{2}\int d^4 x  
\epsilon^{\mu\nu\lambda\sigma} 
\partial_{\nu}B_{\lambda \sigma} \partial_\mu \phi_{\rm ov} \nn\\
&=&\frac{\hbar}{2}\int d^4 x  
\left[\partial_{\mu} (\epsilon^{\mu\nu\lambda\sigma} 
B_{\lambda \sigma} \partial_\nu \phi_{\rm ov}) -B_{\mu \nu} \epsilon^{\mu\nu\lambda\sigma} 
\partial_\lambda \partial_\sigma \phi_{\rm ov}\right]. \nn 
\eea 
We elaborate on these terms by considering a vortex line which is
parallel to the z-axis, whose topological nature can be seen from the term
\bea
\label{eq:circulation0}
\int d^2x 
\,\epsilon^{zt\lambda\sigma} 
\partial_{\lambda}\partial_{\sigma}\phi_{\rm ov}=2\pi. 
\eea 
The topological defects produced by vortex lines can be introduced 
explicitly through
\bea 
\label{eq:circulation1}
\epsilon^{\mu\nu\lambda\sigma} && 
\partial_\lambda \partial_\sigma \phi_{\rm ov}\\
&&=-2\pi \sum_i \int_{\Sigma_i} d\tau d\sigma \epsilon^{\alpha \beta}\partial_{\alpha}X^{\mu}\partial_\beta X^{\nu} \delta^{4}(x^{\mu}-X^{\mu}), \nn
\eea
where $\Sigma_i$ is the worldsheet spanned by the $i$-th 
vortex line, and $\alpha,\, 
\beta \in \{0, 1\}$ label the worldsheet coordinates. $\epsilon^{\alpha \beta}$ is a rank-2 antisymmetric tensor, and we set $\epsilon^{1 0}=1$.
The space
$\sigma \in [\sigma_0, \sigma_1]$, and the time
$\tau$ is chosen to be $t$.  $X^\mu=X^{\mu}(\tau,\sigma)$ 
stands for the coordinates of a vortex line, namely 
$X^{0}=t$, $\mathbf{X}=(x,y,z)$.
Hence, the bulk action of the open vortex line reads 
\bea
S^{\rm v}_{\rm bulk}
&=&-\frac{\hbar}{2}\int d^4 x B_{\mu \nu} \epsilon^{\mu\nu\lambda\sigma} 
\partial_\lambda \partial_\sigma \phi_{\rm ov} \\
&=& \frac{\eta}{2} \sum_i \int_{\Sigma_i} d\tau d\sigma B_{\mu \nu} \epsilon^{\alpha \beta}\partial_{\alpha}X^{\mu}\partial_\beta X^{\nu} , \nn 
\eea
where $\eta\equiv2\pi \hbar$ and $B_{\mu\nu}=B_{\mu\nu}(X^{\mu})$.

Suppose that an ($x$-$y$)-planar dark soliton, 
to which a vortex line can be attached, is located at $z=z_0$. 
We make use of the soliton property of producing a localized $\pi$ phase jump 
 by crossing the soliton plane, i.e. 
 \begin{align}
 \partial_z \phi_s(t,x,y,z)  = \pi \int d^3 \sigma \, 
  \delta^3(x^a-Y^a(\sigma_i))\delta(z-z_0), \nn \\
  \partial_a \phi_s(t,x,y,z)  = \ell \int d^3 \sigma \, 
 \partial_a \hat{\phi}_s(\sigma_i)\, \delta^3(x^a-Y^a(\sigma_i))\delta(z-z_0), 
 \end{align}
 where $\sigma_i$ ($i$ = 1, 2, 3) are the coordinates on the dark 
 soliton plane, $Y^a(\sigma_i) \in \{ t,x,y \}$ is the map from 
 $\sigma_i$ to the spacetime coordinates, $\ell$ is a length scale inserted for 
 dimensional reasons, and $\hat{\phi}_s(\sigma_i)$ is the transverse phase 
 defined only along the dark soliton plane, which in turn contains a smooth 
part and a singular part. In addition, the superfluid current along 
the $z$-direction vanishes at the soliton plane \bea {\rm J}^{z}(t,x,y,z_0)=0, \eea 
which implies $f^{z}(t,x,y,z_0)=0$ and thus imposes a constraint on the 
$B$-field, according to Eq.~(\ref{eq:Bfield}).
Consequently, the boundary term in the action Eq.~\eqref{actionsmooth} vanishes
on the dark soliton surface $-\hbar\int d^3 x f^{z}(z_0,x,y,t)\phi_g=0$. 
As a result, the action for the dark soliton is
\begin{equation}
\label{actionds}
\mathcal{S}_s = \int d^4 x\left(- \hbar f^\mu \partial_\mu \phi_s \right) = 
-\hbar \int d^3 x\left(\ell f^a \partial_a \hat{\phi}_s \right). 
\end{equation}

Combining the boundary term in the action of the open vortex Eq.~\eqref{actionov}:
\bea
\frac{\hbar}{2}\int d^3 x \,
\epsilon^{zabc}B_{bc}\partial_a\phi_{\rm ov}\, ,
\eea
with a,b,c $\in \{ t,x,y \}$, and the dark soliton action Eq.~\eqref{actionds}, 
we obtain the boundary action
\bea
\mathcal{S}^{\rm v}_{\rm boundary} = - \hbar\int d^3 x\left(\ell f^a 
-\frac{1}{2}\epsilon^{abc}B_{bc}\right)\partial_a \hat{\phi}_s. \nn\\
\eea
Here we made the identification $\phi_{\rm ov}(z\rightarrow 
z^{\pm}_0,x,y,t)=\hat{\phi}_s(x,y,t)$, which implies that the 
endpoints of the open vortex lines are the vortex 
excitations living on the soliton surface ($z\rightarrow z^{\pm}_0$). 
As it has previously been done for $\phi_g$, integrating out the smooth 
part of $\hat{\phi}_s$ leads to 
\bea
\partial_a (\ell f^a-\frac{1}{2}\epsilon^{abc}B_{bc})=0\, ,
\eea
whose solution is 
\bea
\label{eq:Ffield}
\ell f^a-\frac{1}{2}\epsilon^{abc}B_{bc}=\frac{1}{2}\epsilon^{abc}F_{bc},
\eea 
where $F_{ab} \equiv 
\partial_a A_b - \partial_b A_a$ is the field strength of $A_a$, and
$A_a$ is a one-form gauge field living on the dark soliton surface
($z\rightarrow z^{\pm}_0$). Therefore, we are left with the 
singular part $\hat{\phi}^{s}_s$ of $\hat{\phi}_s$
\bea
\mathcal{S}^{\rm v}_{\rm boundary} &=& - \hbar\int d^3 x \frac{1}{2}\epsilon^{abc}F_{bc} \partial_a \hat{\phi}^{s}_s \nn\\
&=& -\frac{1}{2} \hbar\int d^3 x \epsilon^{abc}(\partial_b A_c-\partial_c A_b) \partial_a \hat{\phi}^{s}_s \nn\\
&=& -\hbar\int d^3 x A_a (\epsilon^{abc} \partial_b \partial_c \hat{\phi}^{s}_s),
\eea
and for the endpoints of vortex lines, similar to 
Eqs.~\eqref{eq:circulation0}--\eqref{eq:circulation1}, we get  
\be
\epsilon^{abc} \partial_{b}\partial_{c}\hat{\phi}^{s}_s=-2\pi \sum_i \int_{\partial\Sigma_i} d\tau \partial_{\tau}X^{a} \delta^{3}(x^{a}-X^{a}(\tau)),
\ee
where $\partial \Sigma_i$ is the boundary of the worldsheet $\Sigma_i$, and
$X^{a}(\tau)$ are the coordinates of an endpoint of a vortex line, 
namely $X^{0}=t$, $\mathbf{X}=(x,y)$. Then we  obtain 

\bea
\mathcal{S}^{\rm v}_{\rm boundary}
&=& -\hbar\int d^3 x A_a (\epsilon^{abc} \partial_b \partial_c \hat{\phi}^{s}_s) \nn\\
&=& \eta \sum_i \int_{\partial\Sigma_i} d\tau  A_a \partial_{\tau}X^{a},
\eea
where $A^a=A^a(X^a)$.

In terms of the field strengths, the last two terms of the action 
Eq.~\eqref{LagStar} can be written as 
\begin{align}
S_{\rm gauge} & =\frac{m}{2 \rho} \mathbf{f} \cdot \mathbf{f} -\frac{g}{2}(\rho-\rho_0)^2 \nonumber\\
{} & = - \frac{g}{4\ell} \int d^3 x\, 
(\widetilde{F}+\widetilde{B})^2 
- \,\frac{g}{2} \int d^4 x\,\, h_3^2.
\end{align}
Near $\rho \simeq \rho_0$, 
$H_{\mu \nu \lambda}\equiv \partial_{\mu}B_{\nu 
	\lambda}+\partial_{\nu}B_{\lambda\mu}+\partial_{\lambda}B_{\mu\nu}\simeq
H^{0}_{\mu\nu\lambda}+h_{\mu\nu\lambda}$, where $H^{0}_{\mu\nu\lambda}$ 
is the background field with $H^{0}_{123}=\rho_0$, and the fluctuations are 
given by $h_3^2 =h_{\mu\nu\lambda} h^{\mu\nu\lambda} / 6$, through the metric 
$\eta_{\mu\nu} = \textrm{diag} \{-c_s^2,\, 1,\, 1,\, 1\}$ determined by the 
speed of sound $c_s = \sqrt{g \rho_0 / m}$. $\widetilde{F}$ and 
$\widetilde{B}$ are the fluctuations of $F$ and $B$ on the soliton plane 
respectively.

Collecting all the contributions, 
we finally obtain the following dual description of open vortex lines:
\bea
\mathcal{S}^{\ast} &=& S^{\rm v}_{\rm bulk}+S^{\rm v}_{\rm boundary}+S_{\rm gauge} \nn\\
&=& \frac{\eta}{2} \sum_i\int_{\Sigma_i} d \sigma d\tau\, 
B_{\mu\nu} \epsilon^{\alpha\beta} \partial_{\alpha} X^\mu \,\partial_{\beta} X^\nu 
 \nn\\ 
&& +\,\eta \sum_j  \int_{\partial \Sigma_j} d\tau\, A_a \,\partial_\tau X^a -  
\frac{g}{4\ell} \int d^3 x\, 
(\widetilde{F}+\widetilde{B})^2 
\nn\\
&&- \,\frac{g}{2} \int d^4 x\,\, h_3^2\, .
\label{eq:EffStringAction}
\eea
The summation over $\Sigma_i$ includes all the vortex lines and the summation over $\partial 
\Sigma_j$ includes all the endpoints of vortex lines 
attached to dark solitons. 
It is important to note that the resulting action Eq.~(\ref{eq:EffStringAction})
is invariant under the gauge transformations~\cite{Zwiebach}:  
\bea B_{\mu\nu} &&\rightarrow B_{\mu\nu} + \partial_\mu \Lambda_\nu - \partial_\nu 
\Lambda_\mu ,\\ A_a && \rightarrow A_a - \Lambda_a. \eea 

The action given in Eq.~(\ref{eq:EffStringAction}) has two aspects.
First of all, it provides a hydrodynamic 
description of open vortex lines in scalar BECs, which is useful to study the
dynamics of open vortex lines with Dirichlet boundary conditions and vortex-sound interactions. On the other hand, it can be viewed as an effective 
open string action without the tension term, or equivalently 
an action in the 
large $B$-field limit, in the presence of D-branes. Hence, the dual theory
Eq.~\eqref{eq:EffStringAction} might provide a possibility to test some aspects 
of string theory in BEC experiments.

\subsection{Equation of motion}
\label{sect:eom}

In the following we will discuss a simple example of the action
Eq.~\eqref{eq:EffStringAction}. From now on we will 
only consider a single vortex line whose endpoints are attached to two 
parallel dark solitons. Here, for simplicity, we ignore the dynamics of the 
dark solitons and treat them as hard walls. We will also assume that the 
fluctuations of $B_{\mu \nu}$ and $A_{a}$ in space-time are small, and hence the 
field strength parts in $S_{\rm gauge}$ can be neglected since they provide higher 
order terms.

In order to have non-trivial dynamics of a single vortex line, we need to 
introduce in the action Eq.~\eqref{eq:EffStringAction} a phenomenological 
tension term, which plays the role of kinetic 
energy of vortices. This term has been neglected up to now because of the 
assumption of infinitely thin topological defects. Such assumption implies the 
lack of vortex cores and also the absence of a ``physical'' mass of 
the vortex lines, which is responsible, among other dynamical effects, for the 
buoyancy-like forces in inhomogeneous backgrounds \cite{Konotop2004}. As a 
consequence, the tension term can be seen as a remnant of the core structure of 
the vortices in the hydrodynamic limit.

We introduce the tension term in Eq.~\eqref{eq:EffStringAction}
by adding a Polyakov string action \cite{Zwiebach} proportional to the  
the string tension $T$:
\bea
\mathcal{S}^{\rm single}&=&\frac{T}{2 c_s}\int_{\Sigma} d \sigma d\tau \sqrt{-h} 
 \, h^{\alpha\beta} \, \eta_{\mu\nu} \, 
\partial_{\alpha}X^{\mu}\partial_{\beta} X^{\nu}  \nn\\
&&+ \frac{\eta}{2}\int_{\Sigma} d \sigma d\tau\, 
B_{\mu\nu} \epsilon^{\alpha\beta} \partial_{\alpha} X^\mu \partial_{\beta}X^\nu 
\nn\\
 &&+\,\eta \int_{\partial \Sigma} d\tau\, A_a \,\partial_\tau X^a, 
\label{singlevortexline}
\eea
where $h^{\alpha\beta}=\rm diag\{-1/c^2_s,1\}$ is the worldsheet metric, and
$\sqrt{-h}=\sqrt{-{\rm det}(h_{\alpha\beta})}=c_s$. $T$ has the standard dimension $[E] / [L]$ for the string tension and in terms of BEC units $T\sim \pi \hbar^2 \rho_0 /2m$.

Considering a small variation $\delta X^{\mu}$ on top of the background vortex configuration $X^{\mu}$~\cite{bgdfield}, the variation of the  action to the leading order in $\delta X^\mu$ reads
\begin{align}
\delta S^{\rm single} & = \frac{T}{2} \int_\Sigma d\sigma\, d\tau\, 2 \partial_\alpha (\delta X^\mu) \partial^\alpha X_\mu \nn\\
  {} & \quad + \frac{\eta}{2} \int_\Sigma d\sigma\, d\tau\, B_{\mu\nu} \epsilon^{\alpha\beta} (\partial_\alpha \delta X^\mu) \partial_\beta X^\nu \nonumber\\
  {} & \quad + \frac{\eta}{2} \int_\Sigma d\sigma\, d\tau\, B_{\mu\nu} \epsilon^{\alpha\beta} (\partial_\alpha X^\mu) (\partial_\beta \delta X^\nu) \nn\\
  {} & \quad + \frac{\eta}{2} \int_\Sigma d\sigma\, d\tau\, (\partial_\rho B_{\mu\nu}) (\delta X^\rho) \epsilon^{\alpha\beta} \partial_\alpha X^\mu \partial_\beta X^\nu \nonumber\\
  {} & \quad + \eta \int_{\partial \Sigma} d\tau\, (\partial_b A_a) (\delta X^b) \partial_\tau X^a \nn\\
  {} & \quad + \eta \int_{\partial \Sigma} d\tau\, A_a \partial_\tau (\delta X^a) \nonumber\\
  = & \quad  -T \int_{\Sigma} d \sigma d\tau\, 
(\delta X^{\mu})\partial^{\alpha}\partial_{\alpha} X_{\mu}  \nn\\
  {} & \quad + \frac{\eta}{2}\int_{\Sigma} d \sigma d\tau\, 
H_{\mu\lambda\nu} (\delta X^{\mu})  \epsilon^{\alpha\beta} \partial_{\alpha} X^\lambda \partial_{\beta}X^\nu 
\nn\\
  {} & \quad +T \int_{\partial \Sigma} d\tau\, 
(\delta X^{a})\partial_{\sigma} X_{a} \,\nn\\
  {} & \quad +\eta \int_{\partial \Sigma} d\tau\, 
(B_{ab}+F_{ab}) (\delta X^a) \partial_{\tau}X^b. 
\end{align}
Hence the equation of motion of the vortex line in the bulk is given by 
\bea
\label{EOMBULK}
-T \partial_{\alpha}\partial^{\alpha} X_{\mu}+ \frac{\eta}{2} \epsilon^{\alpha \beta} H_{\mu \lambda \nu} \partial_{\alpha} X^{\lambda} \partial_{\beta} X^{\nu}=0, \nn\\
\eea
and the dynamics of the endpoints reads 
\be
\label{EOMB1}
T\, \partial_\sigma X_{a} \big|_{\sigma_1} + \eta\, (B_{ab} + F_{ab}) \partial_{\tau} X^{b} \big|_{\sigma_1} = 0
\ee
and 
\be
\label{EOMB2}
-T\, \partial_\sigma X_{a} \big|_{\sigma_0} - \eta\, (B_{ab} + F_{ab}) \partial_{\tau} X^{b} \big|_{\sigma_0} = 0\, .
\ee

We now consider a special situation. In the bulk $H_{0ij}=0$, and
$H_{ijk}=\rho_0 \epsilon_{ijk}$. On the boundaries $A_0=0$, $\partial_{\tau} 
A_a(\sigma_1)=\partial_{\tau} A_a(\sigma_0)=0$ and 
$B_{nm}(\sigma_{1})=B_{nm}(\sigma_{0})=\rho_0 \epsilon_{nm}$. Here $m, n=1,2$.
For this special case we obtain the following equations 
\bea
0&=&T \left(\frac{\partial^2 X_i}{\partial \sigma^2}-\frac{1}{c^2_s} \frac{\partial^2 X_i}{\partial \tau^2}\right)-\frac{\eta}{2} \epsilon^{\alpha \beta} \rho_0 \epsilon_{ijk} \partial_{\alpha} X^{j} \partial_{\beta} X^{k}\, , \nn\\
0&=& T\, \partial_\sigma X_{n} + \eta\, \rho_0 \epsilon_{nm}\partial_{\tau} X^{m}\, .
\eea

Here we only look for a static solution. For a static vortex line,
\bea
\label{static}
\frac{\partial^2 X_i}{\partial \sigma^2}=0, \quad \partial_{\sigma} X_{n}=0. 
\eea
The general solution would be 
\bea
x=a_1\sigma+a_2 , \quad y=b_1\sigma+b_2, \quad z=c_1 \sigma +c_2 ,
\eea
where $a_1$, $a_2$, $b_1$, $b_2$, $c_1$ and $c_2$ are constants.
By choosing the proper coordinates and taking into account the boundary conditions, the static solution reads
\bea
x=a_2,\quad y=b_2, \quad z(\sigma)=c_1\sigma+c_2,
\eea
which describes a static open vortex line between two dark solitons. Note that this vortex line satisfies Neumann boundary conditions along $x$ and $y$ directions, while fulfills Dirichlet boundary condition along $z$ direction. If we replace the current boundary conditions with periodic boundary conditions, Eq.~\eqref{static} would give a trivial solution, which means that vortex rings must propagate as it is well-known.

In the next two sections we show that such static vortex-soliton composite
topological excitations can indeed be found in the GP theory.
For the sake of comparison with realistic parameters used in current 
experiments of ultracold gases, in what follows we report numerical results for 
systems composed of $^{87}$Rb atoms, with scattering length $a=5.29$ nm, 
confined by harmonic potentials with angular frequencies in the range 
$2\pi\times[10-100]$ Hz.

\section{Open vortex lines}
\label{sect:openvortex}
\subsection{Configurations}
\begin{figure}[tb]
	\flushright
	\includegraphics[width=7.4cm]{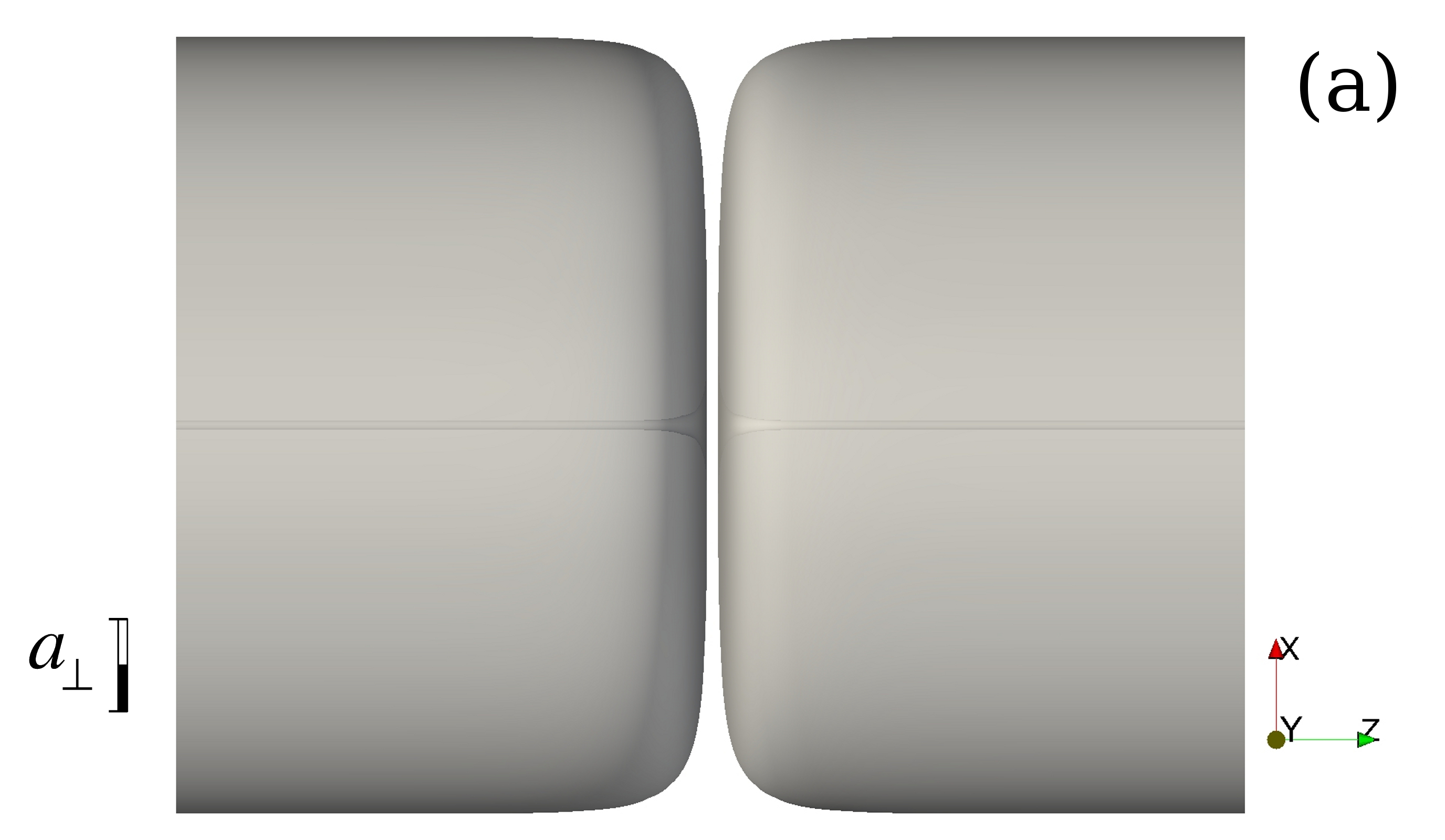}\\
	\includegraphics[width=8.3cm]{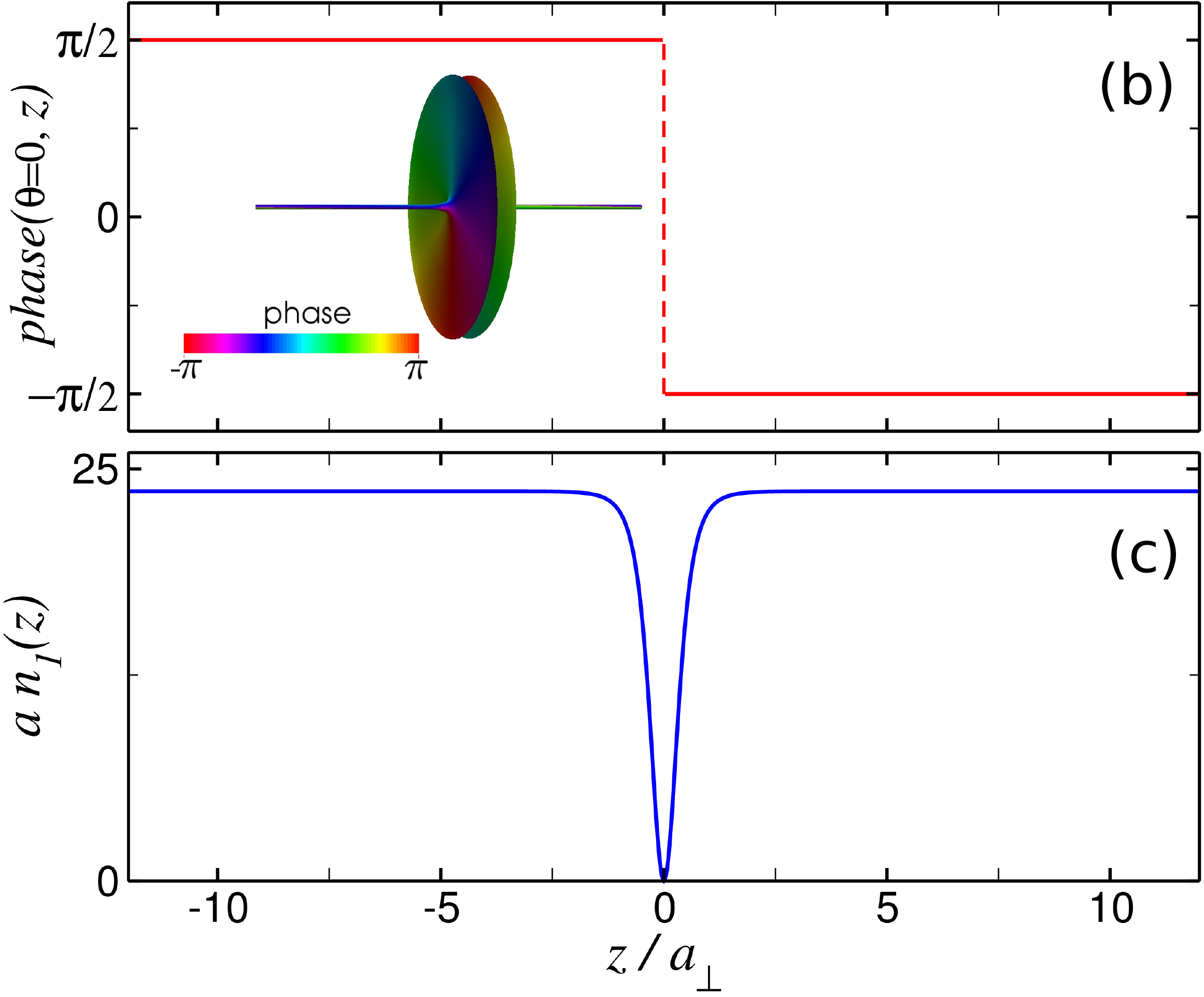}
	\caption{ Singly-charged open vortex lines in a BEC with 
		$\mu=10\,\hbar\omega_\perp$. (a) Semi-transparent density isocontour at 
		5$\%$ of maximum density around the soliton plane. (b) Axial phase for 
		a 
		fixed 
		value of the transverse polar angle $\theta=0$, and density isocontour 
		(inset 
		on the left) of the inner part of the system (capturing the 
		vortex-soliton junction) colored by phase. (c) Dimensionless 
		axial density profile $a\,n_1(z)$.}
	\label{Fig1_os}
\end{figure}

In this section, we consider elongated condensates along the axial 
$z$-direction (assuming $\omega_z=0$) and isotropic trapping 
in the transverse plane, providing the system with a channeled structure. 
In particular, we will consider stationary states containing vortex lines 
attached to solitons. The simplest configuration of this type, containing a single dark soliton, 
is shown in Fig.~\ref{Fig1_os}.
In order to realize this simple configuration we have devised a channeled, cylindrical geometry with an  
axisimetric vortex line:  
\begin{equation}
\psi(z,\theta,r_\perp)=\psi_1(z,r_\perp)e^{i\theta} \, ,
\label{eq:vortex}
\end{equation}
where $\psi_1(z,r_\perp)$ is a real function.
After substituting in Eq.~(\ref{eq:tgpe1}), the vortex solution gives
\begin{align}  
-\frac{\hbar^{2}}{2m}\left(\partial_z^2+\nabla_\perp^2-\frac{1}{r_\perp^2}
\right)\psi_1 + 
V_\perp\psi_1 + g\left\vert \psi_1\right\vert^{2}  \psi_1 
= \mu \psi_1,
\label{eq:vortex_gpe}
\end{align}
where $\nabla_{\perp}^2=\partial_{r_\perp}^2+\partial_{r_\perp}/ r_\perp$ and 
$V_\perp$ accounts for the transverse confinement. This 
is the stationary equation for a vortex state that generates a tangential 
velocity field $\mathbf{v}=\hbar\nabla \arg(\psi)/m$ around the $z$-axis: 
$\mathbf{v}(r_\perp)={\hbar \mathbf{u_\theta}}/{m r_\perp}$,
where $\mathbf{u_\theta}$ is the unit tangent vector.

Next, we search for solutions to the nonlinear Eq.~(\ref{eq:vortex_gpe}) 
including a dark soliton along the axial direction.
An analytical ansatz for this configuration is
\begin{equation}
\psi_1(z,r_\perp)=r_\perp\chi(r_\perp) 
\,\tanh\left(\frac{z}{\xi(r_\perp)}\right) \, ,
\label{eq:openstring}
\end{equation}
where $\xi(r_\perp)=\hbar /\sqrt{m g|\chi(r_\perp)|^2}$ defines a 
radius-varying healing length through the Thomas-Fermi density profile 
$|\chi(r_\perp)|^2=\mu_l(r_\perp)/g$ of a system without the vortex. This 
expression follows the ansatz introduced in Ref. \cite{Mateo2014} for 
3D dark solitons in channeled condensates, and gives a 
good estimate in the strongly interacting regime, where the Thomas-Fermi 
approximation applies. Although the $r_\perp$ factor of 
Eq.~(\ref{eq:openstring}) accounts for the vortex core in a quite simple 
manner, which is characteristic of the corresponding noninteracting system, it 
will turn out to be efficient in getting numerical convergence to the real 
stationary state.

By using the ansatz Eq.~\eqref{eq:openstring} for 
	open vortex lines, we have followed a Newton method to find the exact numerical 
	solution to the full GP Eq.~\eqref{eq:tgpe1}. Fig.~\ref{Fig1_os} shows the 
	features of this configuration (without axial confinement) around a 
	vortex-soliton junction, which leads to Dirichlet boundary conditions for the 
	vortex end points. 
	The panel Fig.~\ref{Fig1_os}(a), presenting a semi-transparent density 
	isocontour of the condensate at 5 $\%$ of maximum density, shows how 
	the presence of the dark soliton breaks the system into two
	phase-separated subsets containing corresponding axisymmetric vortices. These 
	vortices are different entities, as can be seen in the detailed view of 
	panel  Fig.~\ref{Fig1_os}(b). The dark soliton twists their relative 
	phase in $\pi$ radians along the $z$-axis for every value of the azimuthal 
	coordinate 
	$\theta$, and their end points lay aligned on opposite sides of the soliton 
	membrane. Characteristic features of the system are depicted in 
	Fig.~\ref{Fig1_os}(b)-(c): the axial phase jump for a given azimuthal 
	angle, and the axial density of the condensate after integration over the 
	transverse coordinates $n_1(z)=\int |\psi(z,r_\perp,\theta)|^2 r_\perp dr_\perp 
	d\theta$ times the scattering length. The stability of this and more complex 
	configurations within the GP theory will be analyzed in Sections 
	\ref{sect:openvortex}-\ref{sect:halfring}  in realistic condensates.

It is important to note that for scalar BECs, a hypothetical 
configuration with a single 
vortex line attached to a dark soliton is not stationary. For such a case, the 
phase difference between the left and the right side of the dark soliton 
changes continuously along the azimuthal angle from $0$ to $\pi$.  As a 
consequence, the superfluid density can not be zero all along the soliton 
plane, and the superfluid velocity is non-uniform, which makes this a 
transient configuration.

\begin{figure}[tb]
	\includegraphics[width=8.5cm]{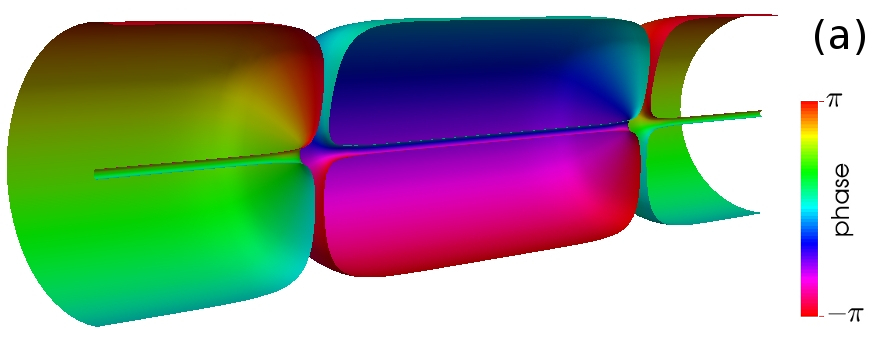}\\
	\includegraphics[width=8.5cm]{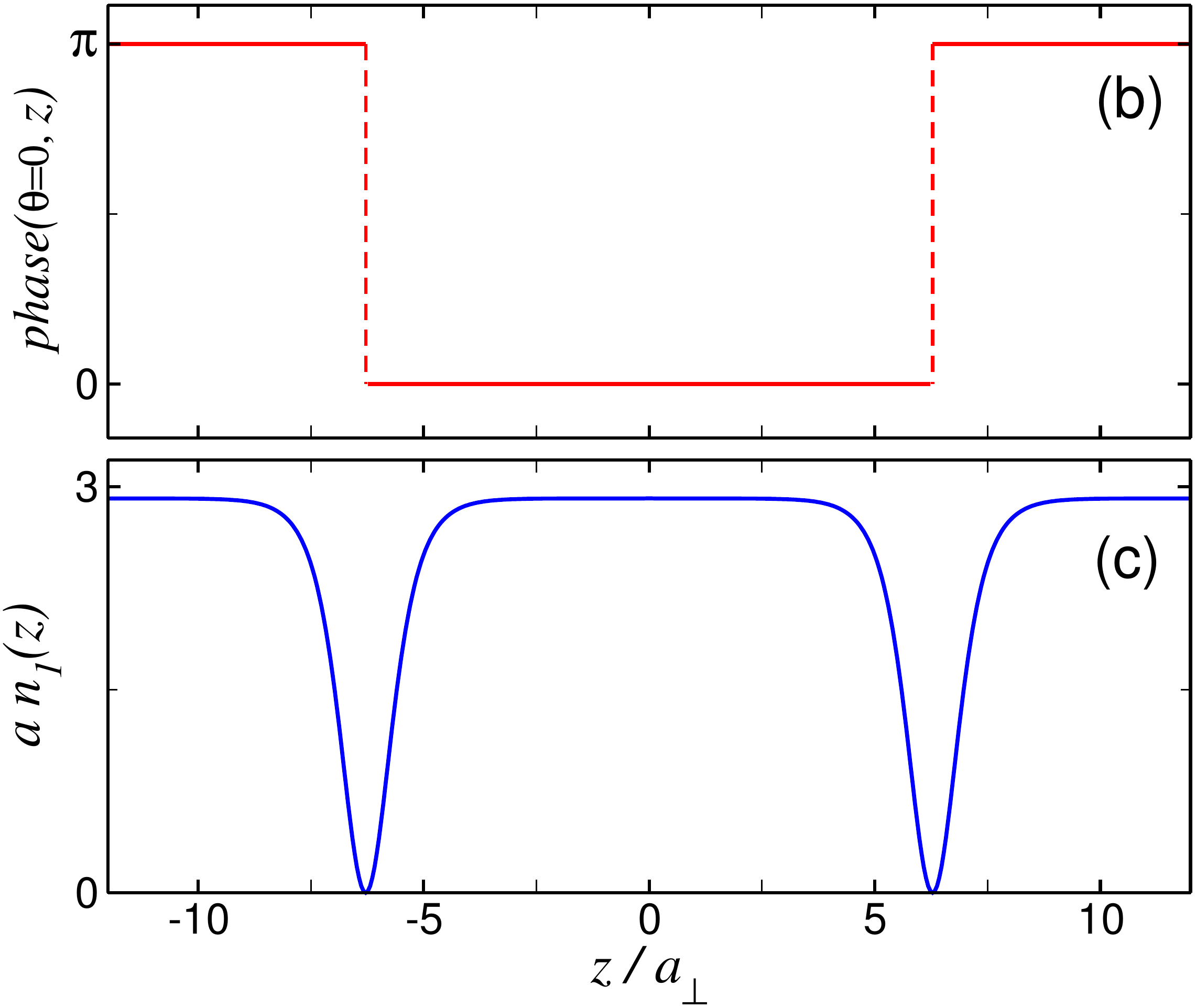}
	\caption{Two dark solitons connected by axisymmetric vortices in a channeled 
		condensate with $\mu=4\,\hbar\omega_\perp$. (a) Perspective view of the
		density isocontour at 5 $\%$ of maximum density, colored by phase, 
		after removing half condensate for better visualization. (b)-(c) Same 
		as in Fig.~\ref{Fig1_os}.}
	\label{Fig2_os}
\end{figure}

It is also interesting to consider configurations containing two nearby 
dark solitons, as the one presented in Fig.~\ref{Fig2_os} for a condensate 
having $\mu=4\,\hbar\omega_\perp$. In this arrangement, the interaction between 
solitons is mediated in the long range by the vortex lines. 
Fig.~\ref{Fig2_os}(a) shows the isocontour at 5 $\%$ of maximum density colored 
according to the complex phase pattern produced by the interplay of solitons 
and vortices. As can be deduced from  Fig.~\ref{Fig2_os}(b)-(c), presenting the 
axial phase and density of the system, the inner region between solitons is 
characterized by a state that differs in the phase from that of 
the outer region.

\subsection{Stability}

States containing open vortex lines, as exemplified by Figs.~\ref{Fig1_os} 
and ~\ref{Fig2_os}, are dynamically stable as long as the dark soliton does 
not decay. As 
it is known, the decay of multidimensional dark solitons is produced by long-wavelength modes excited on the soliton membrane, through the so-called 
snaking instability \cite{Kuznetsov1988}. 
However, such modes can be prevented to appear by means of a tight transverse 
trap, which confines the system to a reduced cross section. In terms of the 
chemical potential, and in the absence of a vortex, a channeled dark soliton is 
stable up to the value $\mu=2.65\,\hbar\omega_\perp$ 
\cite{Mateo2014}. One could 
expect that, since the zero point energy introduced by an axisymmetric vortex 
in the harmonic trap increases in one energy quantum $\hbar\omega_\perp$ 
relative to the ground state, the stability threshold for dark solitons in the 
presence of the vortex would increase by the same amount relative to the case 
without vortex. This argument leads to a threshold localized at
$\mu=3.65\, \hbar\omega_\perp$. As we will see below, it is a reasonable 
estimate, since the unstable frequencies that can be found under such value 
are very small. We have observed that these latter instabilities 
are derived from the junction vortex--soliton (i.e. the boundary conditions imposed 
by the soliton on the vortex ends), and possess slightly different 
amplitudes for different axial lengths of the computational domain considered. 
For particular axial lengths, it is possible to find dynamically stable 
configurations with chemical potentials below the mentioned threshold [as 
the case shown in Fig.~\ref{Fig_stable_OS}(a)], and in the 
general case our results show that the system presents long lifetimes in 
the characteristic units of the trap.

\begin{figure}[tb]
\includegraphics[width=8.5cm]{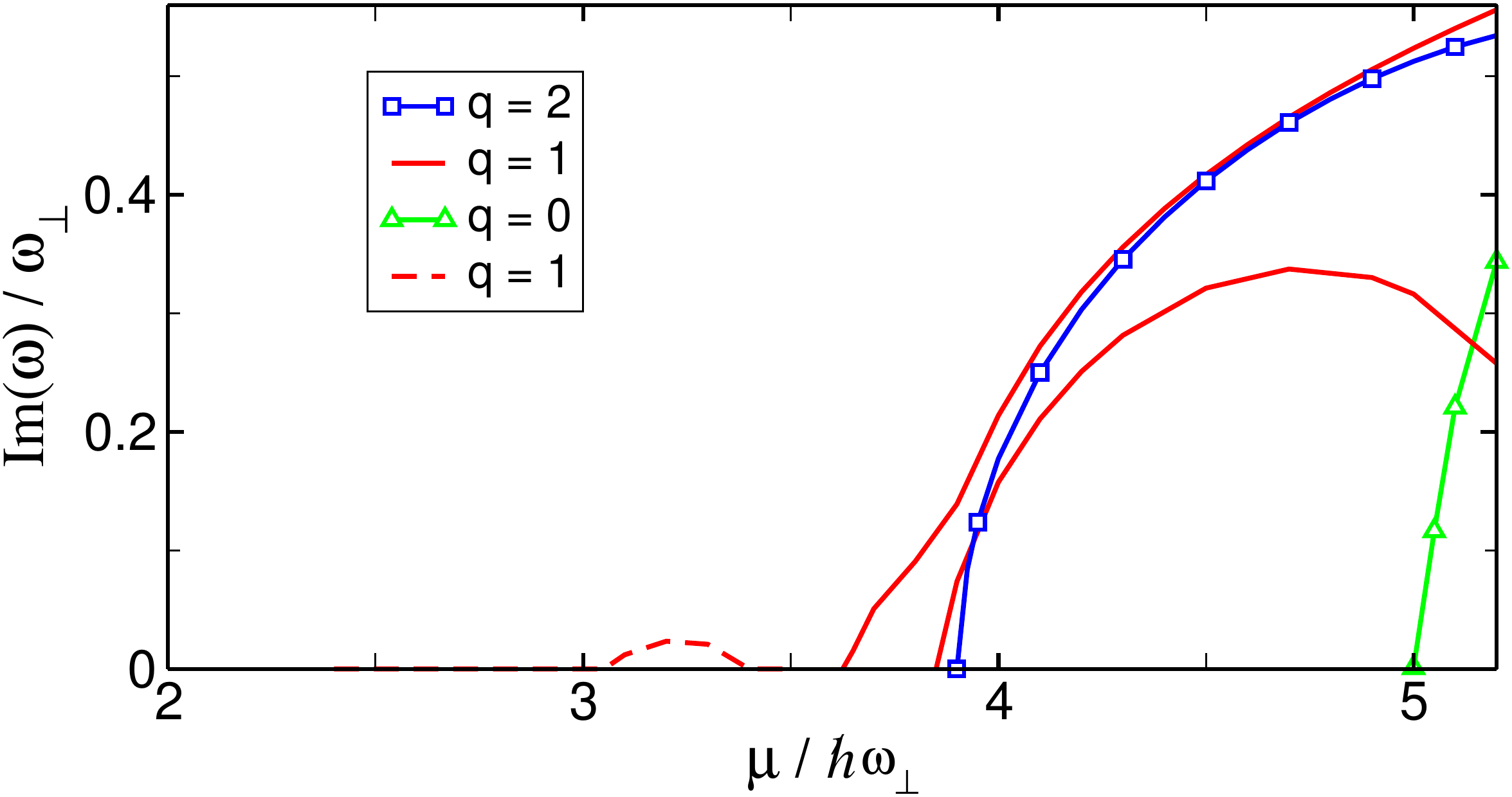}\\
\includegraphics[width=8.5cm]{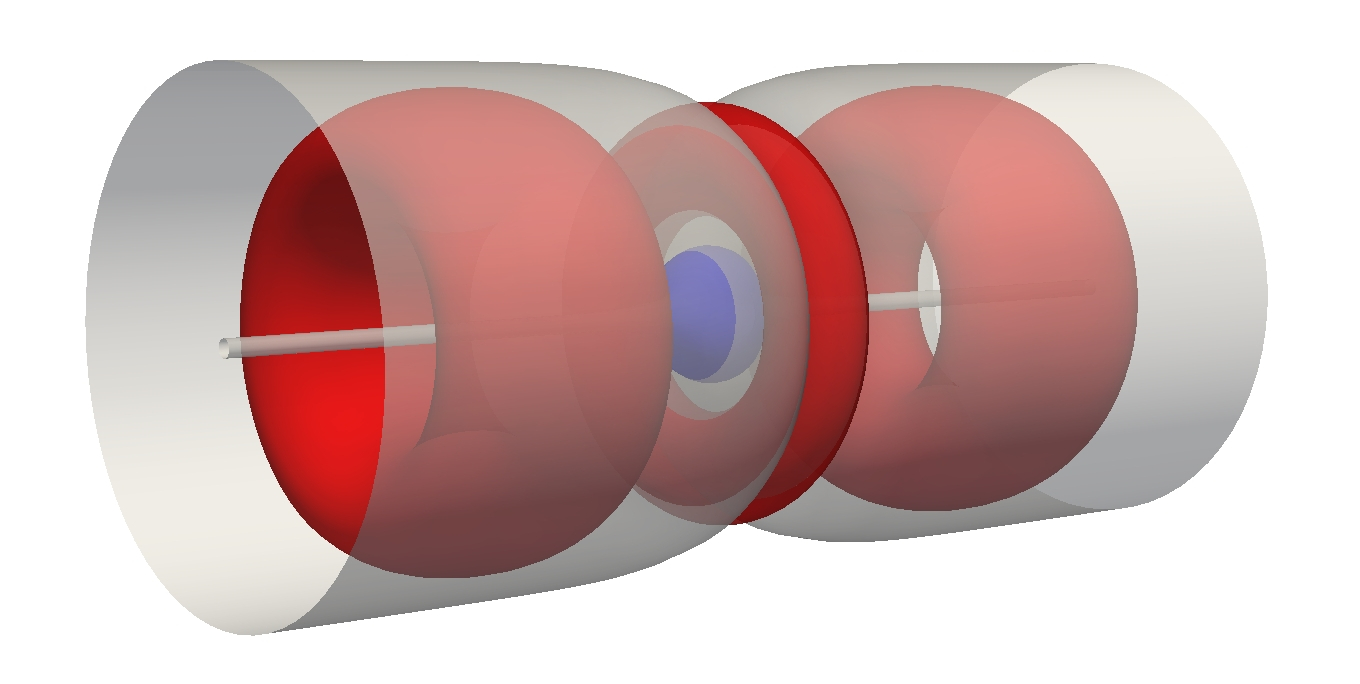}
\caption{Frequencies of unstable modes for open vortex lines as a function of 
the chemical potential (upper panel). The curves correspond 
to the three lowest azimuthal quantum numbers $q$ of the excitation modes (see 
text). The $q=1$ dashed line accounts for excitations derived from the 
vortex-soliton junction, an example of which is given in the lower panel for a 
condensate with $\mu=\,3.2 \hbar\omega_\perp$. The density isocontours (lower 
panel) correspond to the condensate (semi-transparent contour), and to two 
different unstable modes (colour contours) having zero 
(inner blue contour) and $2\hbar$ (outer red toroidal contours) angular 
momentum per 
particle.}
\label{Fig_imag_OS}
\end{figure}

A quantitative analysis of the dynamical stability can be done through the 
Bogoliubov equations (BE) for the linear excitations of the condensate. To this 
aim, we introduce the 
linear modes $\{u(\mathbf{r}),v(\mathbf{r})\}$ with energy 
$\mu\pm\hbar\omega$ to perturb the equilibrium state, i.e. 
$\Psi(\mathbf{r},t)=\psi +\sum_\omega(u \,e^{-i\omega t}+ v^* 
e^{i\omega t})$. After substitution in Eq.~(\ref{eq:tgpe}), and 
keeping terms up to first order in the perturbation, we get
\begin{subequations}
	\begin{align}
	\left( H_L + 2g|\psi|^{2}\right) u +
	g\psi^{2} v = \hbar\omega \, u \, ,
	\\ 
	-g(\psi^*)^2 u -\left( H_L + 2g|\psi|^{2}\right) v
	= \hbar\omega \, v\, ,
	\end{align}
	\label{eq:Bog0} 
\end{subequations}
where $H_L$ is the linear part of the Hamiltonian in Eq.~(\ref{eq:tgpe}), i.e. 
$H_L=-\hbar^{2}\nabla^2/2m+V(\mathbf{r})-\mu$. These equations allow to 
identify the dynamical instabilities of the stationary state $\psi$, which are 
associated to the existence of $\omega$ frequencies with non-vanishing 
imaginary parts.

For the vortex state $\psi(z,\theta,r_\perp)=\psi_1(z,r_\perp)e^{i\theta}$ the 
BE Eq.~\eqref{eq:Bog0} read
\begin{subequations}
\begin{align}
\left( H_L + 2g\psi_1^{2}\right) u +
g\psi_1^{2}e^{i2\theta} v = \hbar\omega \, u \; ,
\\ 
-g{\psi_1}^2e^{-i2\theta} u -\left( H_L + 2g\psi_1^{2}\right) v
 = \hbar\omega \, v\; .
\end{align}
\label{eq:vortex_Bog} 
\end{subequations}
We search for the modes of the functional form 
$\{u(z,r_\perp)e^{i(q+1)\theta},v(z,r_\perp)e^{i(q-1)\theta}\}$ with 
$q=0,\pm 1,\pm 2,...$. After adding and substracting both Eqs. 
(\ref{eq:vortex_Bog}), we achieve
\begin{align}
\left( H_1+\frac{\hbar^2 (q^2+1)}{2mr_\perp^2} + g_\pm \psi_1^{2} \right) f_\pm 
= (\hbar\omega- \frac{\hbar^2 q}{mr_\perp^2}) f_\mp \; ,
\label{eq:vortex_Bog1}
\end{align}
where $H_1=-\hbar^2(\partial_z^2+\partial_{r_\perp}^2+\partial_{r_\perp}/ 
r_\perp)/2m+V_\perp-\mu$, $g_\pm=(2\pm 1)g$, and 
$f_\pm(z,r_\perp)=u(z,r_\perp)\pm 
v(z,r_\perp)$.

\begin{figure}[tb]
\includegraphics[width=8.5cm]{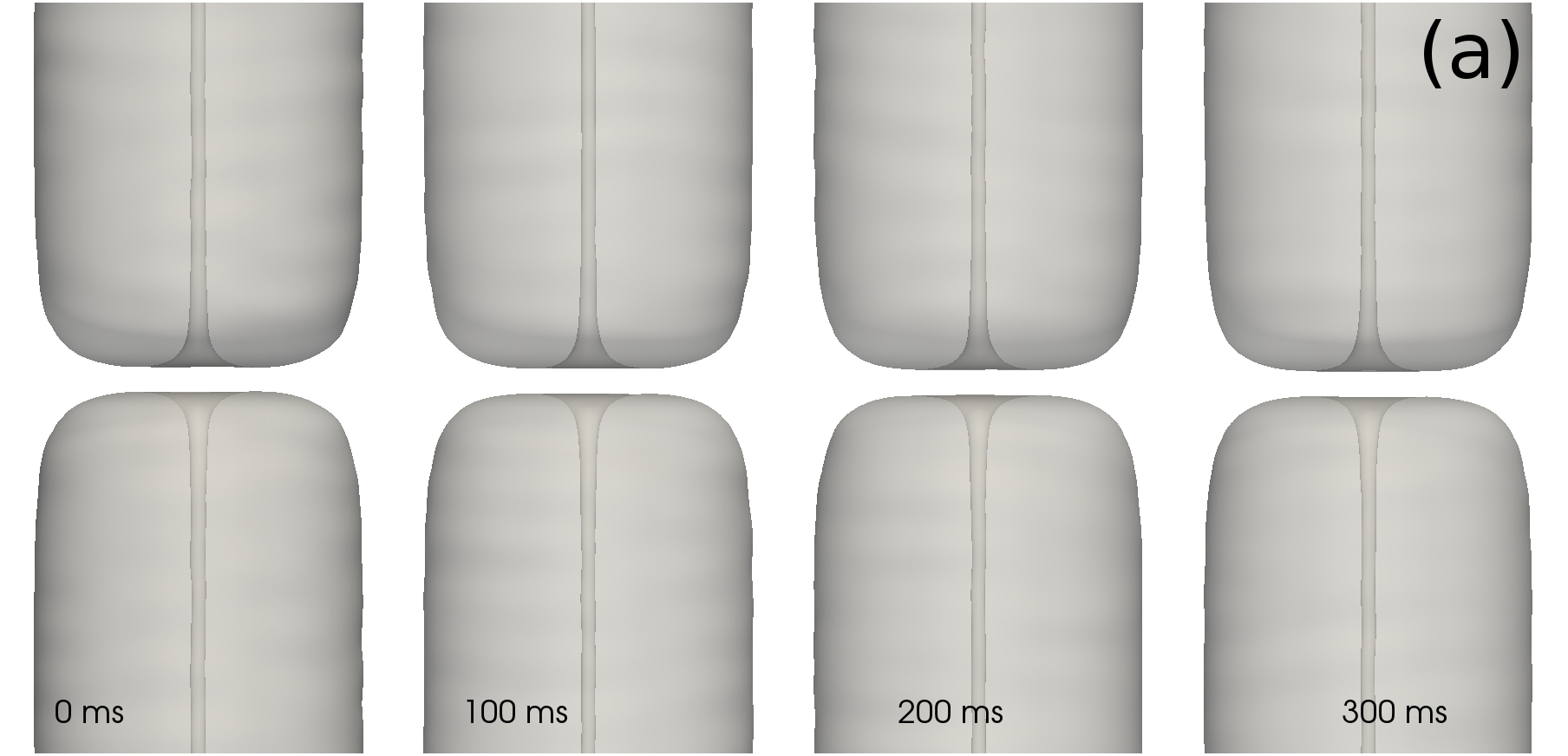}\\
\vspace*{0.25cm}
\includegraphics[width=8.75cm]{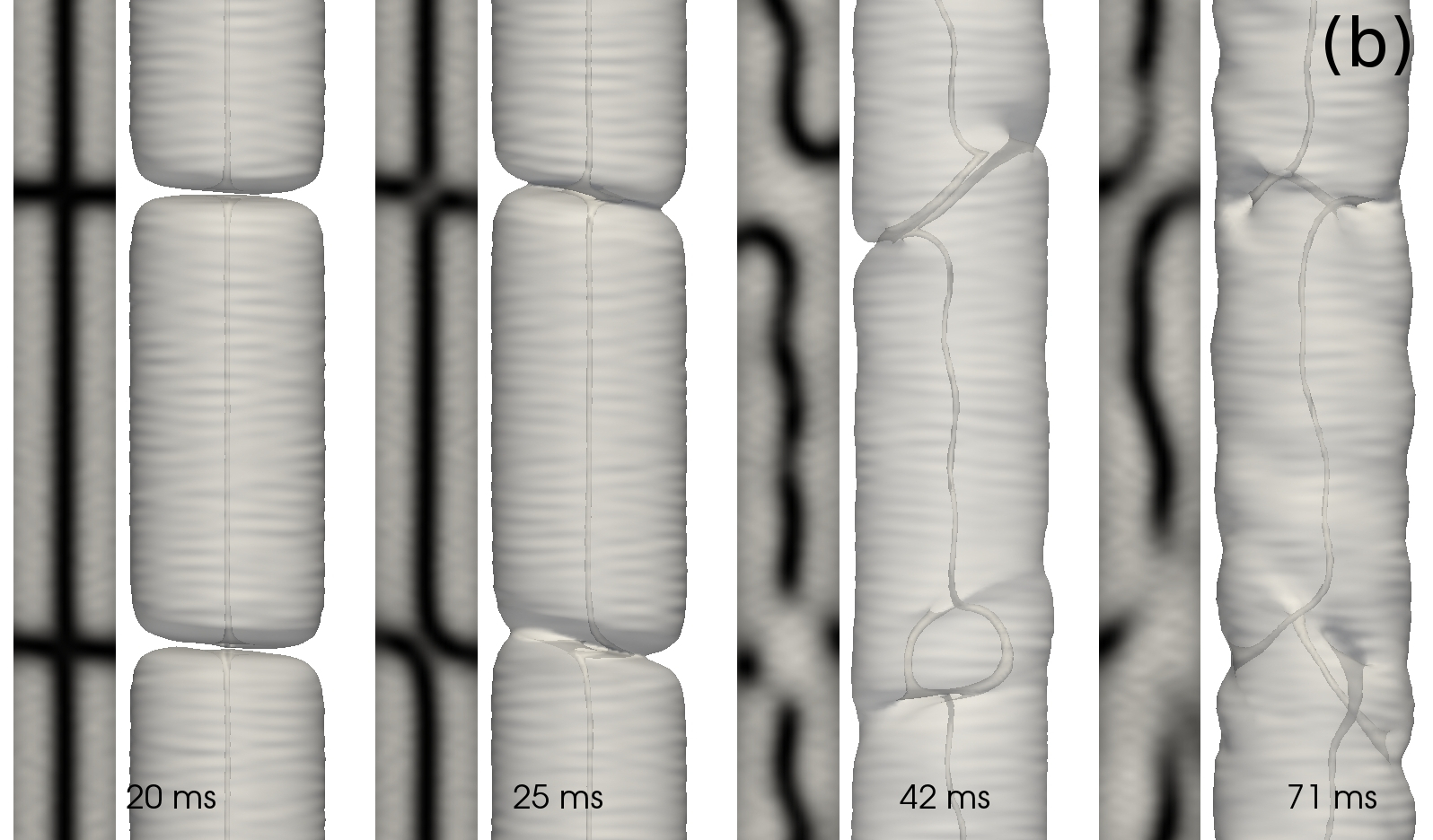}
\caption{Real time evolution of open vortex lines after adding random 
perturbations to the corresponding stationary states. (a) Semi-transparent 
density isocontours at 5 $\%$ of maximum density around the soliton 
position for a stable state with $\mu=3.0\, \hbar\omega_\perp$ and 
$\omega_\perp/2\pi=100$ Hz. (b) Same as (a) for an unstable case 
containing two solitons, with  $\mu=5.1\,\hbar\omega_\perp$ and 
$\omega_\perp/2\pi=71.3$ Hz, showing a decay dynamics dominated at early time 
by the $q=1$ mode (see text). For every time, 
along with the density isocontour of the whole system on the right, a narrow 
slice along the $z$-axis shows the density in a greyscale, with vortices in 
black, on the left. }
\label{Fig_stable_OS}
\end{figure}
Bifurcations from the dark soliton state occur if 
Eqs.~\eqref{eq:vortex_Bog1} have non-trivial solutions for 
$\omega=0$~\cite{Brand2002,Mateo2015}. In this case, and for $q=0$ 
modes, Eqs. 
(\ref{eq:vortex_Bog1}) are linear Schr\"{o}dinger equations for $f_\pm$, with 
effective potentials given by $V_\pm=V_\perp+\hbar^2/2mr_\perp^2+g_\pm 
\psi_1^{2}$. 
In particular, the equation for $f_-$ is identical to the GP equation, thus 
admitting the solution $f_-^G=\psi_1(z,r_\perp)$, apart from a global phase. 
This solution is the Goldstone mode associated to the breaking of the 
continuous symmetry of the phase. Since $f_-^G$ presents an axial node (the one 
of the soliton), there must be another solution to Eq.~(\ref{eq:vortex_Bog1}) 
without axial nodes, and then with lower axial energy. This energy difference, 
associated to the axial degrees of freedom, can be released for the excitation
of transverse modes in the condensate, which can produce the decay of the 
dark soliton. Following a procedure parallel to that used in
Ref.~\cite{Mateo2014}, based in a separable ansatz for $f_\pm$ within the 
Thomas-Fermi regime, the bifurcation points for $q=0$ can be estimated to
appear at chemical potential values $\mu_0 =\sqrt{2}\,({2p + 2}) \,\hbar 
\omega_\perp$, where $p=1,2,\dots$ is a radial quantum number.  
The pair $(p,q)$ characterizes the corresponding transverse unstable 
modes localized at the soliton plane, and indicates the number of radial and 
azimuthal nodal points, respectively. Specifically, for $(p=1,q=0)$ the 
predicted unstable mode will appear at $\mu=5.66 \,\hbar\omega_\perp$, which is 
close to the value ($\approx 5$) found by numerically solving 
Eq.~(\ref{eq:vortex_Bog1}).

The upper panel of Figure~\ref{Fig_imag_OS} shows the numerical solutions 
to Eq.~(\ref{eq:vortex_Bog1}) for the unstable 
excitation frequencies of open-vortex-line states. As can be seen, the unstable 
modes with $q=1,2$ ($p=0$) appear before those with $q=0$. The latter 
excitations introduce radial nodes ($p\neq0$) in the soliton 
plane, whose energy cost is higher 
than the kinetic energy excess ($\propto \hbar^2 q^2/2m$) of the azimuthal 
nodes generated by the modes with $q=1,2$. In particular, for a computational 
domain with periodic boundary conditions and axial length $4\pi\,a_\perp$, the 
excitation of transverse modes with $q=1$ above $\mu=3.0\,\hbar\omega_\perp$ 
marks the threshold for instability. As previously commented, the small bump 
extended between $3.05\,\hbar\omega_\perp$ and $3.4\,\hbar\omega_\perp$ on the
$\mu$ axis is due to small instabilities derived from the vortex--soliton 
junction, and are represented by the ($q=1$)--dashed 
line of Figure~\ref{Fig_imag_OS}. To illustrate this point, 
the lower panel of Figure~\ref{Fig_imag_OS} depicts the density isocontours of 
an open vortex state with $\mu=\,3.2\,\hbar \omega$ (semi-transparent contour) 
and its only unstable modes coming from the vortex--soliton junction (colored 
contours). Such unstable modes are not exclusively localized around the 
junction. This 
specific instability ceases to act for an intermediate range of chemical 
potentials, where the snaking instability makes its appearance through the 
$q=1$ solid curve.

We have cross-checked our results obtained from the linear stability analysis, 
by evolving in real time the stationary states $\psi$ with different chemical 
potentials, after adding a random Gaussian perturbation 
$\psi\rightarrow\psi+\delta\psi$. The upper panel of Fig.~\ref{Fig_stable_OS} 
presents an example of such dynamics for a condensate with 
$\mu=3\,\hbar\omega_\perp$, just below the instability threshold. The system 
remains nearly unaltered during the whole evolution, thus stable according to 
the linear analysis of Fig.~\ref{Fig_imag_OS}; only smooth oscillations due to 
sound waves can be observed.

For states with two solitons, as shown in Fig.\ref{Fig2_os}, whenever the 
distance between solitons is large in comparison with the healing length, the 
stability analysis follows that of a single
soliton. To illustrate their characteristic dynamics, we have selected an 
unstable system with $\mu=5.1\,\hbar\omega_\perp$, shown in the lower panel of 
Fig.~\ref{Fig_stable_OS}. As can be seen, at about 20 ms, the snaking instability 
starts to operate at the planes of the solitons, while the vortices begin 
to lose their alignment. At later time, two new vortices appear as a remainder 
of the two initial solitons, along with manifest oscillations of the initial 
straight axisymmetric vortices.

\section{Half vortex ring states}
\label{sect:halfring}

\begin{figure}[t]
\includegraphics[width=8.5cm]{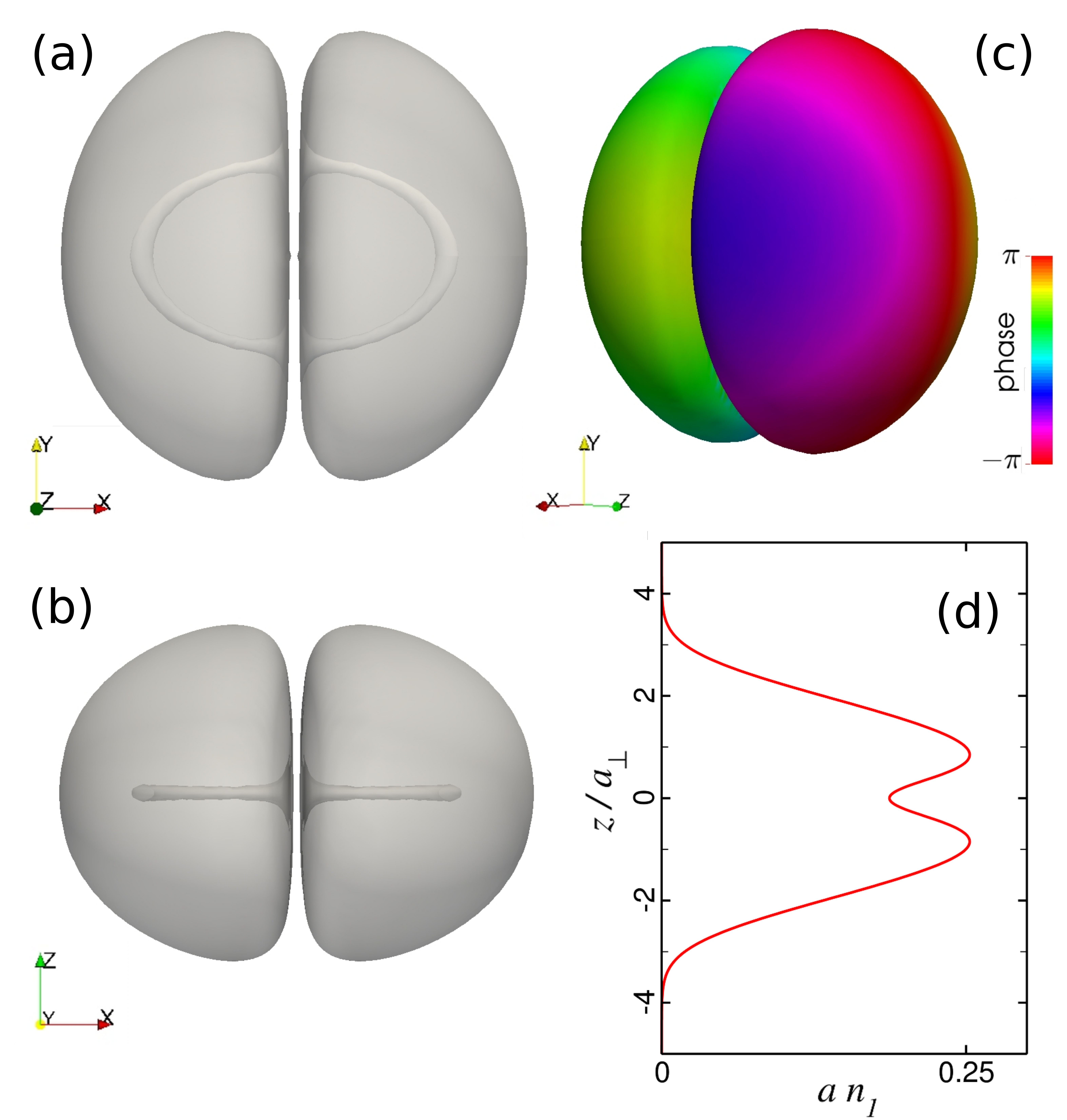}
\caption{Half vortex rings attached to a dark soliton in an oblate 
condensate with $\mu=6\,\hbar\omega_\perp$ confined by a harmonic trap with 
aspect ratio $\lambda=1.4$. Density isocontours at 5$\%$ of maximum density are 
shown for the semi-transparent top and side views (left panels) and the 
perspective view colored by phase (top right). (d) Column density profile 
along the $z$-axis. }
\label{Fig3_os}
\end{figure}
In this section we consider a variant layout with curved open vortex 
lines having both ends attached to the same 
planar dark soliton. The resulting structures bear some resemblance to the 
U-shaped vortices found in elongated systems \cite{Rosenbusch2002}, where the 
vortex lines bend near their end points. On the 
contrary, here we focus on half vortex rings. These objects have been 
studied in the realm of classical fluids \cite{Akhmetov2009}, and can propagate 
long distances attached to the water surface without decaying 
(see for instance~\cite{Hu2005}, where they are shown to provide the mechanism 
of motion for light insects). This configuration mimics an open string whose
both ends are attached to the same D-brane, which is another basic configuration
of an open string.

\begin{figure}[t]
\includegraphics[width=8.5cm]{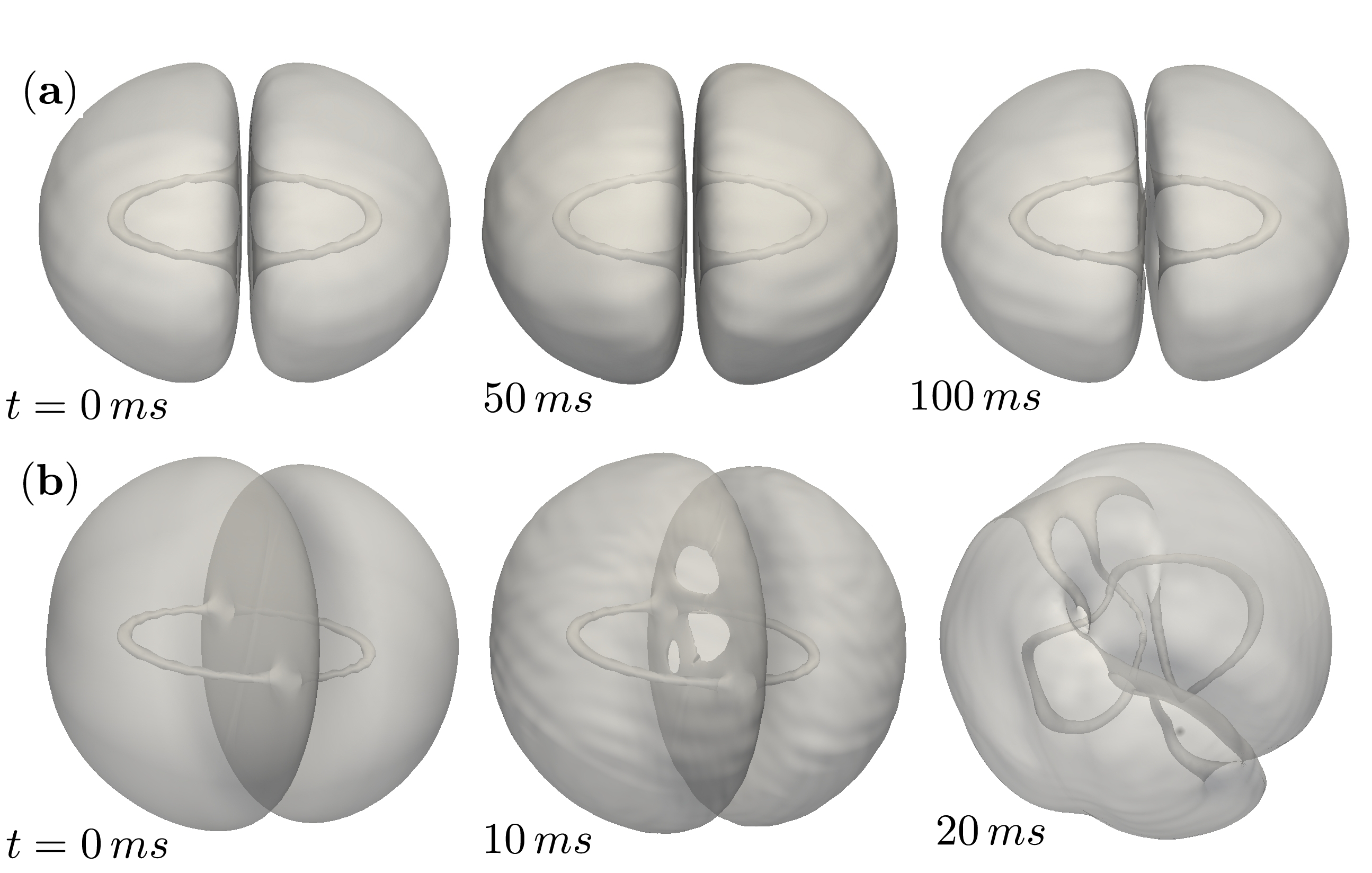}
\caption{Real time evolution, after adding a random Gaussian perturbation, of 
half vortex rings in oblate BECs. Semitransparent density isocontours 
at 5 $\%$ of maximum density are shown at different times for condensates 
with: (a) $\lambda=1.4$, $\mu=6\,\hbar\omega_z$, and $\omega_z/2\pi=11.5$ Hz; 
(b) $\lambda=1.1$, $\mu=8.9\,\hbar\omega_z$, and $\omega_z/2\pi=57.5$ Hz.}
\label{Fig4_os}
\end{figure}

In order to generate half vortex rings in a BEC, we proceed with two 
steps. We first search for stable vortex rings, which are later split 
into two halves by imprinting a planar dark soliton. To this end, we 
have chosen near-spherical, harmonically trapped condensates, where 
the conditions for the stability of vortex rings have been 
analytically predicted for aspect ratios in 
the range $1\leq\lambda\leq 2$ within the strongly interacting 
regime \cite{Horng2006}, and have been numerically 
corroborated in some particular cases near the weakly interacting regime 
\cite{Bisset2015}. Our results show that the analytical prediction breaks down 
for small values of the 
chemical potential, where, on the other hand, the conditions for the 
stability of solitons can be expected. As a result, we have not found a 
stable half vortex ring, although it is possible to observe robust 
configurations with weak instabilities. This is the case presented
in Fig.~\ref{Fig3_os}, corresponding to the aspect ratio $\lambda=1.4$ and 
the chemical potential $\mu= 6\,\hbar\omega_z$. 
The translucent density isocontours Fig.~\ref{Fig3_os}(a)-(b)
show the condensate from 
two perpendicular views, and the colored 
density isocontour Fig.~\ref{Fig3_os}(c) reproduces the resulting phase pattern 
from the interplay between the soliton (lying across the $x$-coordinate)  and 
the vortices. By integration of the density over the $x$-coordinate [panel 
Fig.~\ref{Fig3_os}(d)], it 
is possible to observe the density depletion produced by the vortices at $z=0$. 
As demonstrated in Fig.~\ref{Fig4_os}(a), showing the real time evolution of 
this state obtained from the numerical solution of the full GP equation 
(\ref{eq:tgpe}), the half vortex ring is a robust state, which is able to 
survive under perturbations during $100$ ms in a harmonic trap with 
$\omega_z/2\pi= 11.5$ Hz.

Other choices of parameters can lead to different lifetimes for 
half vortex rings. As an alternative 
example, in Fig.~\ref{Fig4_os}(b) we present a near-spherical condensate with 
$\lambda=1.1$ and a higher chemical potential $\mu= 8.9\,\hbar\omega_z$, in a 
trap with $\omega_z/2\pi= 57.5$ Hz. In this case, the 
dark soliton decay begins around 10 ms, and the later evolution shows 
complex dynamics. As a general trend, for high values of $\mu$ we have observed a 
chaotic scenario at the later stages of the half-vortex-ring decay, including 
the appearance of new vortex lines that split or reconnect.

\section{Discussion and Conclusions}
\label{sect:conclusions}

In this paper two variants of open vortex lines in scalar BECs have been reported, 
having either one end or two ends of a vortex attached to a planar soliton. 
In the latter case, consisting of half vortex rings, we have shown that these 
configurations present long lifetimes for typical parameters of current 
experiments. For channeled condensates containing straight vortices attached 
to solitons, we have demonstrated the existence of dynamically stable 
states and identified the bifurcation points of the first unstable 
modes. In both variants, a necessary condition for the corresponding 
configurations to be stationary requires that the vortex lines appear in pairs, 
as shown in 
Figs.~\ref{Fig1_os},~\ref{Fig2_os},~\ref{Fig_stable_OS},~\ref{Fig3_os}.

The stationary configurations we considered in this paper are either stable 
(vortex lines attached to planar dark solitons) or long-lived (U-shaped vortex 
lines), which indicate their feasibility for BEC experimental realizations. To 
this aim, the well-established 
experimental techniques in BECs developed for the observation of topological 
defects, such as optical phase imprinting \cite{Burger1999,Denschlag2000} or laser 
stirring \cite{Madison2000}, can be used. We propose here two possible ways for 
elongated condensates
adequately devised to prevent the snaking instability \cite{Weller2008}. By 
means of phase imprinting, both a transverse planar soliton and a 
pinned straight vortex along the $z$-axis could be simultaneously seeded. 
Alternatively, the laser stirring of the atomic cloud could be applied to 
drive the condensate into rotation around the $z$-axis, so that an 
energetically stable, singly charged vortex can be generated. After this, a
planar soliton could be imaged onto the condensate by phase imprinting, in
order to split the initial vortex into two vortex lines attached to the 
soliton.
With regard to the realization of half 
vortex rings, the experimental procedures are more elaborate, since they would 
involve the controlled generation of vortex rings \cite{Anderson2001}, within a 
regime of dynamical stability, followed by the phase imprinting of a soliton.
These settings can open up a way to advance in the fairly unexplored domain of 
the interplay between vortices and solitons.

To summarize, we have demonstrated that stationary, robust states composed 
of vortex lines attached to planar dark solitons can be found in scalar 
BECs. Among them, we have reported on dynamically stable configurations in 
elongated systems for small values of the chemical potential.
Our results follow from the Gross-Pitaevskii theory 
applied to realistic condensates of ultracold gases, and allow the study of
vortex lines with Dirichlet boundary conditions. In addition, we have shown 
that in the hydrodynamic limit the dynamics of open vortex lines can be well characterized by the dual description of the Gross-Pitaevskii theory, which can be viewed as a (3+1)-dimensional effective string theory.
This connection might pave the way to test some analytical predictions of 
string theory with experimental realizations in ultracold gases. To advance in 
this way, the further analysis of solutions to the 
equation of motion derived from the string functional
Eq.~(\ref{eq:EffStringAction}) and the comparison with numerical simulations 
from the Gross-Pitaevskii equation are required, which will be reported 
elsewhere \cite{BECstring3}.

\vspace{0.5cm}
\section*{Acknowledgments}
We would like to thank S. Giaccari, J. Gomis, V. Pestun and S. Gubser for 
many useful discussions. We are also very grateful to J. Brand, A. Bradley 
and B. P. Blakie for reading previous versions of this manuscript.

\bibliographystyle{unsrt}
\bibliography{BEC1}

%

\end{document}